\documentclass[fleqn,usenatbib]{mnras}

\usepackage[T1]{fontenc}

\DeclareRobustCommand{\VAN}[3]{#2}
\let\VANthebibliography\thebibliography
\def\thebibliography{\DeclareRobustCommand{\VAN}[3]{##3}\VANthebibliography}

\usepackage{graphicx}	% Including figure files
\usepackage{amsmath}	% Advanced maths commands
\usepackage{amssymb}

\title[The Bayesian Global Sky Model (B-GSM)]{The Bayesian Global Sky Model (B-GSM): A Calibrated Low Frequency Sky Model for EoR Applications}   

\author[G. Carter]{
George Carter,$^{1}$\thanks{E-mail: gtc30@cam.ac.uk}
Will Handley,$^{1}$
Mark Ashdown,$^{2}$
and Nima Razavi-Ghods$^{2}$
\\
$^{1}$Institute of Astronomy, University of Cambridge, Cambridge, United Kingdom\\
$^{2}$Cavendish Laboratory, Department of Physics, University of Cambridge, Cambridge, United Kingdom\\
}

\date{Accepted XXX. Received YYY; in original form ZZZ}

\pubyear{\the\year{}}

\emergencystretch=\maxdimen
\usepackage{caption}
\begin{document}
\label{firstpage}
\pagerange{\pageref{firstpage}--\pageref{lastpage}}
\maketitle

% Abstract of the paper
\begin{abstract}  
We present the Bayesian Global Sky Model (B-GSM), a new absolutely calibrated model of the diffuse Galactic foreground at frequencies $\leq$408 MHz. We assemble a dataset of publicly available diffuse emission maps at frequencies between 45 MHz and 408 MHz, along with absolute temperature data from the EDGES radiometer between 40 and 200 MHz. We use nested sampling to perform a joint Bayesian analysis of these two datasets and determine posterior distributions of: spatially resolved diffuse components, spectral parameters for the diffuse emission, and calibration corrections for each observed diffuse emission map. Using Bayesian model comparison, we find that the low-frequency sky is optimally modelled by two emission components, each following a curved power-law spectrum. The spectrum for the first component has a spectral index of $\beta_1=-2.633\pm0.002$ and a curvature of $\gamma_1=0.014\pm0.001$,  while the second has $\beta_2=-2.108\pm0.008$ and $\gamma_2=-0.424\pm0.008$.  The diffuse maps require temperature-scale corrections of 1\% to 29\%, and zero-level adjustments of a few kelvin to a few hundred kelvin. We find that the Haslam 408~MHz map is well calibrated, requiring a scale correction of $1.029\pm0.003$ ($\sim3\%$ adjustment) and zero-level correction of $0.91\pm0.05$ kelvin. Posterior predictions for the sky’s absolute temperature are in excellent agreement with EDGES data, indicating accurate calibration. The posterior sky predictions agree with observations within statistical uncertainty across all frequencies. However, agreement varies by position, with the largest discrepancies in the Galactic plane. This is the second paper in the B-GSM series, the low-frequency sky model (as well as all code and data) is available for download.

\end{abstract}

% Select between one and six entries from the list of approved keywords.
% Don't make up new ones.

%%%%%%%%%%%%%%%%%%%%%%%%%%%%%%%%%%%%%%%%%%%%%%%%%%

%%%%%%%%%%%%%%%%% BODY OF PAPER %%%%%%%%%%%%%%%%%%

\section{Introduction}
Detection of the cosmological 21cm signal is limited by contamination by bright foreground emission that exceeds the expected signal by 3-6 orders of magnitude~\citep{Pritchard2012,LFSM}. This foreground emission is dominated by diffuse Galactic synchrotron radiation, with additional contributions from Galactic free-free emission \citep{Lian2020} and extragalactic radio sources. To identify and extract the cosmological 21cm signal, it is critical to accurately model and remove this foreground contamination. As such, the lack of an accurate calibrated low-frequency foreground model, compounded by the shortage of modern large area low-frequency diffuse emission surveys, poses a major challenge for 21cm cosmology.

Previous sky models, such as the Global Sky Model (GSM)~\citep{GSM}, its 2016 update ~\citep{Zheng2016}, and the Low Frequency Sky Model~\citep{LFSM} perform component separation using a Principal Component Analysis (PCA) of the diffuse emission survey maps that form their datasets. These models, while widely used, have notable limitations for low-frequency applications. They neglect the variability and uncertainty in the calibration of the underlying surveys, which is known to be significant \citep{Monsalve2021,Spinelli2021}, and these sky models are primarily based on high frequency data from well above 1~GHz. These high-frequency datasets may not be representative of low-frequency foregrounds. Additionally, the PCA based component separation used by these models does not provide any estimate for the uncertainty on the predicted sky. These issues are particularly relevant to cosmic dawn and reionisation studies, where the relevant frequency range is below 200~MHz~\citep{Liu2013,Pritchard2012}.

In this paper, we present a new data driven low-frequency sky model, the Bayesian Global Sky Model (B-GSM). B-GSM is based on a Bayesian analysis of two independent datasets. The first of these is a set of ten publicly available diffuse emission survey maps, which are spatially resolved but potentially poorly calibrated. The second is a set of absolute temperature measurements from the EDGES radiometer~\citep{Monsalve2021,Mozdzen2016,Mozdzen2018}, which has limited spatial information but is well calibrated. Conditioning B-GSMs posterior on both datasets allows us to produce a spatially resolved sky model, that is also absolutely calibrated.

We use the Bayesian simultaneous component separation and calibration algorithm introduced in the first B-GSM paper~\citep{Carter2025} to generate samples from the joint posterior distribution of; the spatially resolved diffuse emission components, spectral parameters, and calibration corrections for the survey maps. Nested sampling \citep{Skilling2004} is used to compute Bayesian evidence, allowing us to determine posterior distributions for emission components and perform rigorous Bayesian model comparison \citep{Trotta2008} to select the optimal model parameterization. Please see the first paper in the series~\citep{Carter2025} for a full discussion of B-GSMs approach to calibration and component separation.

Our novel Bayesian approach allows us to determine the full posterior distribution of the predicted sky, inherently quantifying uncertainty in our predictions. By conditioning the joint posterior on both (poorly calibrated) spatially resolved diffuse emission surveys and (well calibrated, but not spatially resolved) EDGES data, we are able to ensure absolute calibration for our posterior predictions. In this way B-GSM addresses the limitations of previous sky models; ensuring robust 
uncertainty quantification, absolute calibration, and nonarbitrary model parametrisation that is guided by the dataset.

The remainder of this paper is structured as follows. In section~\ref{s:the dataset} we present a summary of publicly available low-frequency diffuse emission surveys, and select a dataset of diffuse maps. Additionally, we briefly discuss the calibration issues present in our diffuse dataset, and introduce an independent absolute temperature dataset from the EDGES radiometer~\citep{Monsalve2021,Mozdzen2016,Mozdzen2018}. In section~\ref{s:brief theory}, we provide a brief overview of our simultaneous component separation and calibration algorithm. Sections~\ref{s:Bayesian evidence} and \ref{s:highest evidence posterior} present the results of Bayesian model comparison, and discuss the posterior for our the highest evidence model. 
Finally, in section~\ref{s:conclusions} we present our conclusions and discuss future research directions.

\begin{figure}
    \centering
    \includegraphics[width=0.95\linewidth]{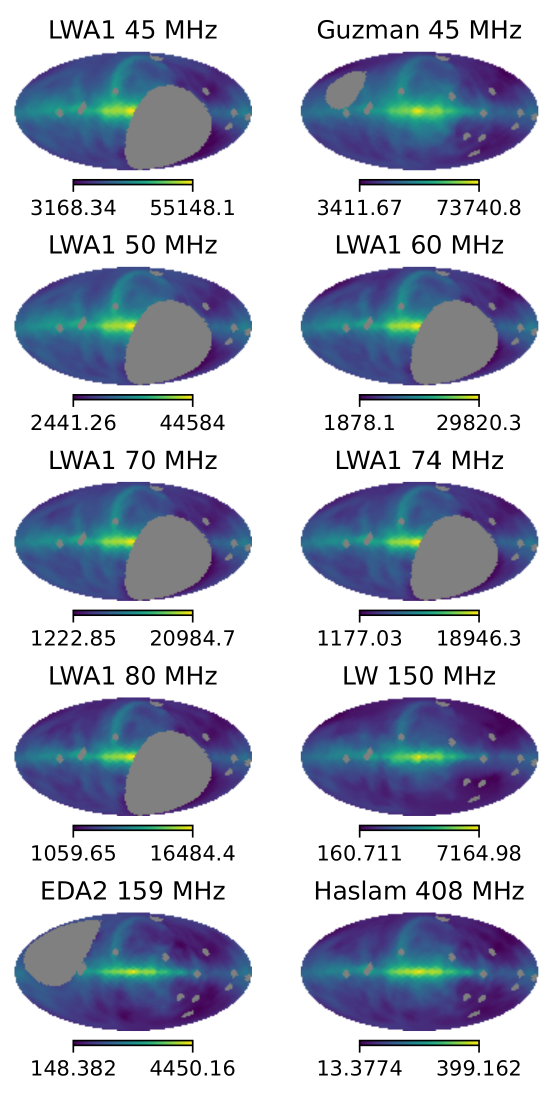}
    \caption{The diffuse maps used in B-GSMs diffuse dataset. Maps are shown in mollweide projection on a log scale in units of kelvin.}
    \label{fig:diffuse_dataset}
    \includegraphics[width=0.94\linewidth]{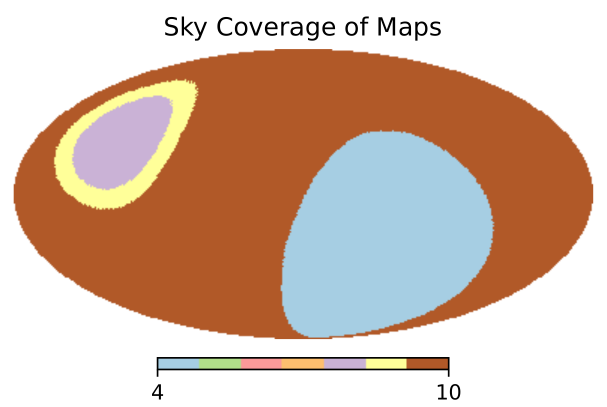}
    \caption{The sky coverage of maps in our diffuse dataset, we see that all regions have at least 4 observations.}
    \label{fig:coverage}
\end{figure}
 
\section{B-GSM Dataset}\label{s:the dataset}

\begin{table*}
    %\begin{center}
    %\caption{Available total power radio surveys.}
    \centering
    \caption{A List of Available Diffuse Emission Surveys in the range $20\ \mathrm{MHz}$ to $1\ \mathrm{GHz}$.}
    \label{t:dataset table}
    \begin{tabular}{ccccc}
      Survey & $v$ (MHz) & Coverage (Declination) (\degr) & Resolution & Reference(s)\\\hline
        DARO & 22 & $-28\degr, +80\degr$ & $1.2\degr \times 1.7\degr$ & \cite{Roger1999DARO}\\
        LWA1 & 35 & $-40\degr,+90\degr$  & $4.8\degr \times 4.5\degr$ & \cite{LFSM}\\
        OVRO-LWA & 36.528 & $-30\degr,+90\degr$  & $26.0\arcmin$ & \cite{OVRO-LWA}\\
        LWA1 & 38 & $-40\degr,+90\degr$  & $4.5\degr \times 4.1\degr$ & \cite{LFSM}\\
        LWA1 & 40 & $-40\degr,+90\degr$  & $4.3\degr \times 3.9\degr$ & \cite{LFSM}\\
        OVRO-LWA & 41.760 &$-30\degr,+90\degr$ & $23.3\arcmin$ & \cite{OVRO-LWA}\\
        LWA1 & 45 & $-40\degr,+90\degr$ & $3.8\degr \times 3.5\degr$ &\cite{LFSM}\\
        Guzman & 45 & $-90\degr, +70\degr$ & $3.6\degr$ & \cite{Guzman2010}\\
        OVRO-LWA & 46.992 & $-30\degr,+90\degr$  & $20.9\arcmin$ & \cite{OVRO-LWA}\\
        LWA1 & 50 & $-40\degr,+90\degr$  & $3.4\degr \times 3.1\degr$ & \cite{LFSM}\\
        OVRO-LWA & 52.224 & $-30\degr,+90\degr$  & $18.7\arcmin$ & \cite{OVRO-LWA}\\
        OVRO-LWA & 57.456 & $-30\degr,+90\degr$  & $18.0\arcmin$ & \cite{OVRO-LWA}\\
        LWA1 & 60 & $-40\degr,+90\degr$  & $2.8\degr \times 2.6\degr$ & \cite{LFSM}\\
        OVRO-LWA & 62.688 & $-30\degr,+90\degr$  & $17.8\arcmin$ & \cite{OVRO-LWA}\\
        OVRO-LWA & 67.920 & $-30\degr,+90\degr$  & $17.6\arcmin$ & \cite{OVRO-LWA}\\
        LWA1 & 70 & $-40\degr,+90\degr$ & $2.4\degr \times 2.2\degr$ & \cite{LFSM}\\
        OVRO-LWA & 73.152 & $-30\degr,+90\degr$  & $18.6\arcmin$ & \cite{OVRO-LWA}\\
        LWA1 & 74 & $-40\degr,+90\degr$  & $2.3\degr \times 2.1\degr$ & \cite{LFSM}\\
        LWA1 & 80 & $-40\degr,+90\degr$  & $2.1\degr \times 2.0\degr$ & \cite{LFSM}\\
        LW 150~MHz (All-Sky) & 150 & All Sky & $5\degr$ & \cite{LW150}\\
        EDA2 & 159 & $-90\degr, +60\degr$ & $3.1\degr$ & \cite{EDA2}\\
        Haslam & 408 & All Sky & $56\arcmin$ & \cite{Remazeilles2015}\\
        Dwingeloo & 820 & $-7\degr, +85\degr$ & $1.2\degr$ & \cite{Dwingloo}\\
        \\\hline
        \vspace{0.3cm}
    \end{tabular}
    %\end{center}
\end{table*}
\subsection{Diffuse Dataset}
To construct our low-frequency sky model we must assemble a dataset of large area diffuse emission survey maps covering the frequency range of interest. We began by performing a literature review of publicly available sky surveys covering the range $20\ \mathrm{MHz}$ to $1.0\ \mathrm{GHz}$. The results of this search are summarised in table \ref{t:dataset table}. All diffuse maps included in the table are publicly available for download from \href{https://lambda.gsfc.nasa.gov/}{LAMBDA} with the exception of the Guzman 45~MHz map, available from \href{https://vizier.cds.unistra.fr/viz-bin/VizieR-4}{VizieR}. 

We note that, many of the sky maps have large beam sizes and that the available maps have very limited coverage of the sky south of declination $-30\degr$. Several experiments aiming to observe the Epoch of Reionization (EoR) and the Cosmic Dawn, e.g the HERA \citep{HERA} and EDGES \citep{Bowman2018}, are located in the Southern Hemisphere. This lack of low-frequency southern sky observations presents a significant problem when constructing a foreground model for EoR applications. 

\newpage
Due to the lack of southern sky observations, we include all maps with southern sky coverage (regardless of their resolution) in our diffuse dataset. Our dataset includes; the Guzman 45~MHz map \citep{Guzman2010}, the Landecker-Wielebinski (LW) 150~MHz all sky map \citep{LW150}, the engineering development array 2 (EDA2) 159~MHz \citep{EDA2}, the all sky Haslam 408~MHz map \citep{Remazeilles2015}, and the LWA1 45, 50, 60, 70, 74, and 80~MHz maps \citep{LFSM}. This gives us a diffuse dataset of ten maps at 45, 50, 60, 70, 74, 80, 150, 159, and 408~MHz. We have chosen to omit any sky surveys above 408~MHz, this is due to B-GSM being focused on modelling the low frequency sky for EoR and 21-cm cosmology applications.

The ten diffuse emission maps, used for this study, are then pre-processed. For all ten diffuse maps we smooth the map to have a FWHM of $5\degr$ (the beam size of the lowest resolution map), mask out the 14 brightest point sources, and subtract the contribution from the CMBR ${T_\mathrm{CMBR}=2.7260\pm0.0013}$ kelvin~\citep{Fixsen_2009}. Additionally, we apply the calibration corrections, found by Monsalve et al. 2021~\citep{Monsalve2021}, to the Guzman 45~MHz and LW 150~MHz maps. At this stage, we do not attempt to calibrate the other maps in our dataset. The final pre-processed diffuse dataset is shown in figure~\ref{fig:diffuse_dataset}. The sky coverage of this dataset is summarised by the map in figure \ref{fig:coverage} which shows the number of observed frequencies for each region of the sky. We can see that all regions of the sky have at least 4 observations, and that for all regions of the sky the dataset covers the full frequency range 45-408~MHz.

Uncertainty maps are only available for a limited number of the maps in our dataset. The LWA1 maps at 45, 50, 60, 70, 74, and 80~MHz have published uncertainty maps \citep{LFSM}. For the Guzman 45~MHz and the LW 150~MHz maps approximate uncertainty maps were published by Monsalve et al. 2021 \citep{Monsalve2021} based on their re-calibration for these two maps. The Haslam 408~MHz map does not have a published uncertainty map. For B-GSM we will use the published uncertainty maps at 45, 50, 60, 70, 74, 80, and 150~MHz, and we will assume an uncertainty of 10\% for each pixel in the map at the frequencies where we do not have published uncertainty maps.
\subsection{Absolute Temperature Dataset}
The observed maps that form our diffuse dataset are known to have inconsistent and inaccurate calibration for both their temperature-scale and temperature zero-level (see \cite{Monsalve2021,Spinelli2021}). In B-GSM we want to address this calibration uncertainty, and ensure that the posterior sky predictions are absolutely calibrated. To achieve this we introduce a second independent absolute temperature dataset. This second dataset will act as a ground truth, allowing us to infer calibration corrections for the diffuse dataset and ensure absolute calibration of the posterior predicted sky. 

For our independent absolute temperature dataset, we will use measurements from the EDGES experiment \citep{Bowman2018}. As the EDGES dataset has not been publicly released, we did not have access to the raw antenna temperature measurements. Instead, in this work, we make use of the EDGES low-band and high-band spectral index measurements \citep{Mozdzen2016,Mozdzen2018} and the EDGES antenna temperature measurements at the reference frequencies 75~MHz~\citep{Mozdzen2018} and 150~MHz~\citep{Monsalve2021}.   

\begin{figure}
    \centering
    \includegraphics[width=0.99\linewidth]{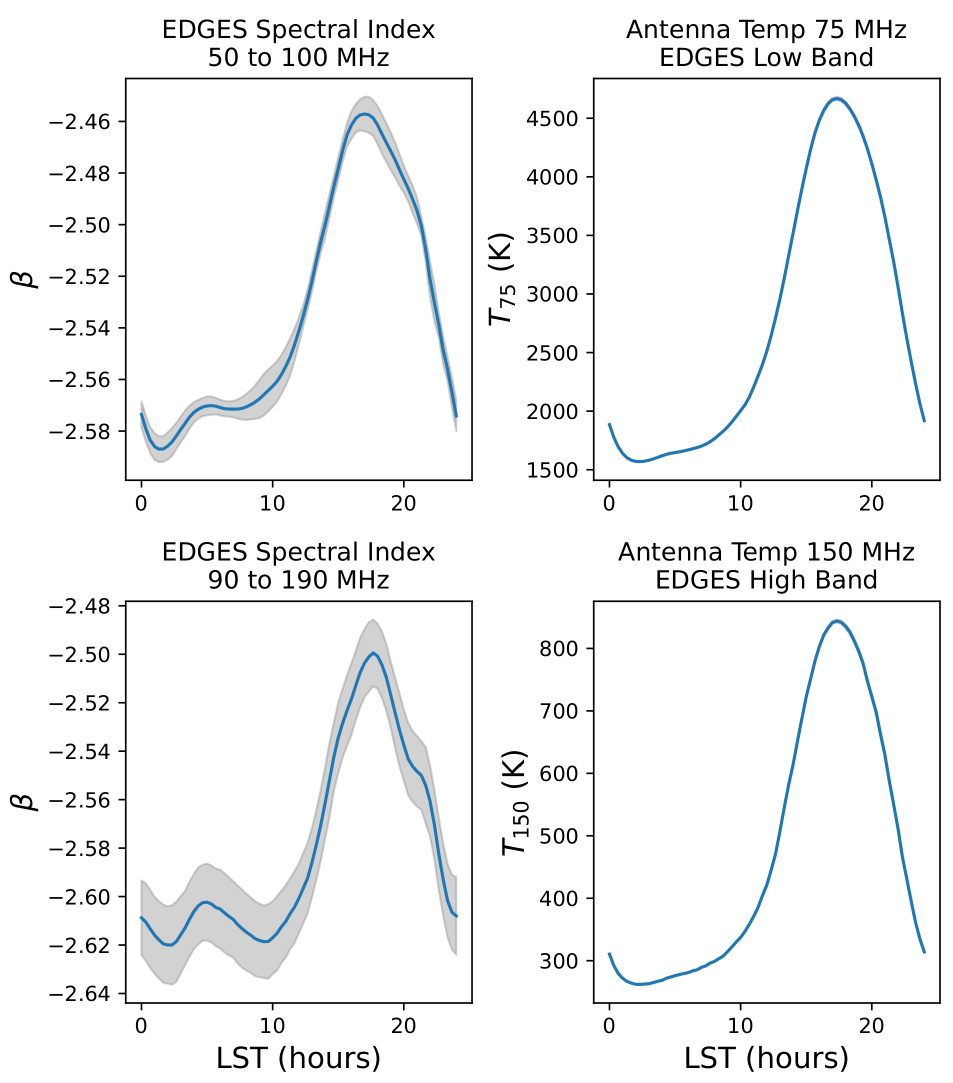}
    \caption{Spectral index and antenna temperature as a function of LST. The top row shows EDGES low band observations of the spectral index between 50 and 100~MHz and antenna temperature at 75~MHz, data from \citep{Mozdzen2018}. The bottom row shows the EDGES high band observations of the spectral index between 90 and 190~MHz and antenna temperature at 150~MHz, spectral index data from \citep{Mozdzen2016} antenna temperature data from \citep{Monsalve2021}.}
    \label{fig:EDGES_data}
\end{figure}

In figure~\ref{fig:EDGES_data} we show these spectral indexes and antenna temperature measurements.  The top row shows the spectral index measurements from the EDGES low band system, covering the frequency range 50-100~MHz and the full 24 hours of LST, along with the antenna temperature at 75~MHz (data taken from \cite{Mozdzen2018}). The bottom row shows the spectral index measurements from the EDGES high band system~\citep{Mozdzen2016}, covering the frequency range 90-190~MHz and the full 24 hours of LST, along with the antenna temperature at the reference frequency 150~MHz~\citep{Monsalve2021}. The blue line in each subplot shows the reported measurements, and the greyed region covers the $1\sigma$ uncertainty.

The low band spectral indexes and 75~MHz antenna temperatures are used to generate $T$ vs LST (antenna temperature as a function of LST) curves at 5~MHz spacings between 40~MHz and 100~MHz. The high band spectral indexes and 150~MHz antenna temperatures are used to generate $T$ vs LST curves at 5~MHz spacings between 105~MHz and 200~MHz. We compute antenna temperatures every 20 minutes of LST between 0 hours and 24 hours, giving us a dataset of sky temperature values taken at every 20 minutes of LST at 5~MHz spacings between 40 and 200~MHz. Note that the EDGES low-band and high-band spectral indexes used in this study are chromaticity corrected. A future research goal is to re-release B-GSM based on an analysis of the raw EDGES antenna data.

\newpage
\section{B-GSM Theory (A Brief Overview)}\label{s:brief theory}
\begin{figure}
    \centering
    \includegraphics[width=0.95\linewidth]{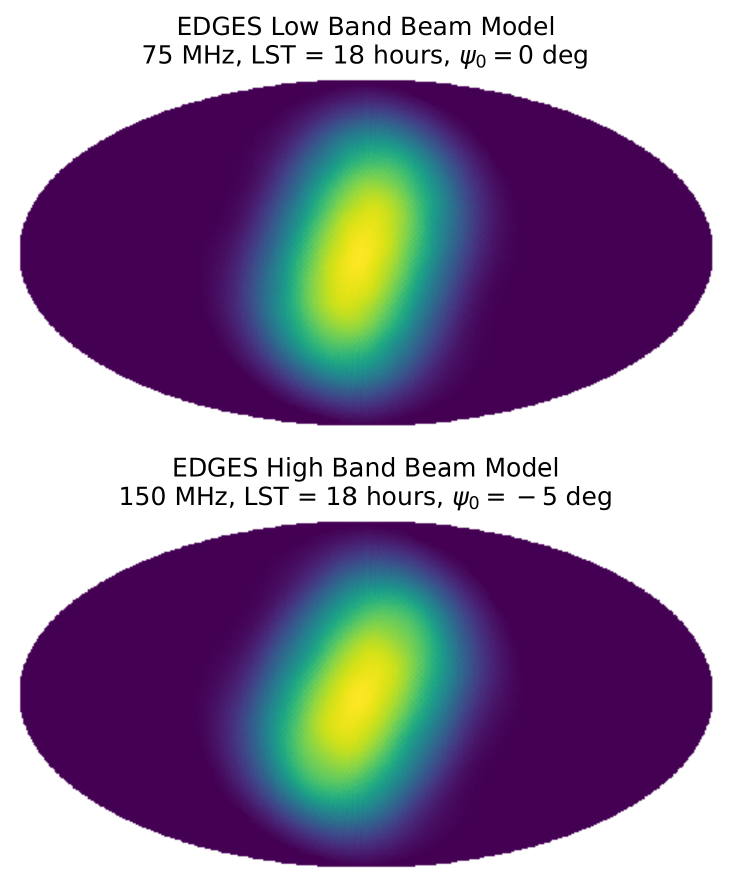}
    \caption{EDGES low band and high band beam models shown in Galactic coordinates for observations taken at latitude $-26.7\deg$ \citep{Mozdzen2018} at Local Sidereal Time (LST) 18 hours. The term $\psi_0$ is the azimuth angle of the dipole excitation axis relative to North at the time of observations, $\psi_0$ values from \citep{Mozdzen2018,Monsalve2021}. Low band beam model from \citep{Mahesh:2021rly} high band beam model (Mahesh et al. Private Communication).}
    \label{fig:beams}
\end{figure}
The simultaneous component separation and calibration algorithm for B-GSM is detailed in full in the first paper of this series~\citep{Carter2025}, we will briefly summarise the algorithm in this section. As with previous sky models e.g. the Global Sky Model (GSM)~\citep{GSM} we begin by assuming the true underlying sky is described as a sum of $k$ emission components. Each with a spatial amplitude map, $M_c(\Omega)$, and a spectrum, $S^c(v)$:
\begin{equation}\label{eq:component_sep}
    D_\mathrm{true}(\Omega,v) = \sum_{c=1}^{k}{M_c(\Omega)S^c(v)}.
\end{equation}

The observed sky at frequency $v$ will have calibration errors and thermal noise. As in \citep{Monsalve2021} we assume that the correction to the calibration for each diffuse map takes the form of a global zero level shift and a global scale factor correction. Specifically, we assume that (for frequency $v$) the correctly calibrated map, $D_{v,\mathrm{cal}}(\Omega)$, is related to the observed map in our dataset, $D_v(\Omega)$, by:
\begin{equation}\label{calibration eq}
    D_{v,\mathrm{cal}}(\Omega) = a_v D_v(\Omega) + b_v.
\end{equation}
where, $a_v$ is the correction to the temperature scale and $b_v$ is the correction to the temperature zero level for the map at frequency $v$. It is these correctly calibrated observed maps that can be related to our true underlying signal. Accounting for the noise in the observed maps, denoted as \( N(\Omega,v) \), and noting that the correction to the temperature scale also affects the noise, we find that:
\begin{equation}\label{eq:cal_obs_maps}
a_v D_v(\Omega) + b_v = \left(\sum_{c=1}^{k}{M_c(\Omega)S^c(v)}\right)  + a_v N(\Omega,v),
\end{equation}

Our aim is to determine the joint posterior distribution of the set of component maps, $\Vec{M}$, parameters of the spectral model, $S$, and calibration corrections for each of the diffuse maps in our dataset, $a_v$ \& $b_v$, conditioned on both the dataset of diffuse maps, $D$, and the EDGES absolute temperature dataset, $E$:
\begin{equation}
    P(a,\Vec{b},\Vec{M},S|E,D) = \frac{P(E,D| a,\Vec{b},\Vec{M},S) P(a,\Vec{b},\Vec{M},S)}{P(E,D)}.
\end{equation}
From the product rule for conditional probability, and the independence of the datasets, the likelihood function can be written as:
\begin{equation}
    P(E,D|a,\Vec{b},\Vec{M},S) = P(E|\Vec{M},S)P(D|a,\Vec{b},\Vec{M},S). 
\end{equation}

The term $P(D|a,\Vec{b},\Vec{M},S)$ is the likelihood of observing the diffuse dataset for a specific set of component maps $\Vec{M}$ and spectral model $S$. This is a Gaussian likelihood given by:
\begin{multline}
    2\ln P(D|a,\Vec{b},\Vec{M},S) = \sum_p-\left[\Vec{d}_p - a^{-1}\left(S\Vec{M}_p -\Vec{b}\right)\right]^T N_p^{-1}\\ \left[\Vec{d}_p - a^{-1}\left(S\Vec{M}_p -\Vec{b}\right)\right] - \ln\left(\left|2\pi N_p \right|\right).
\end{multline}
The sum runs over the pixels in the diffuse maps, we have assumed uncorrelated noise between pixels and frequencies.

The term $P(E|\Vec{M},S)$ is the likelihood of observing the EDGES absolute temperature dataset for a specific set of component amplitude maps $\Vec{M}$ and spectral model $S$. This is a Gaussian likelihood given by: 
\begin{multline}
    \label{EDGES likelihood term}
    2\ln{P\left(E|\Vec{M},S\right)} = -\sum_{\mathrm{LST},v} \left(\frac{T_{E,v,\mathrm{LST}} - T_{\mathrm{mod},v,\mathrm{LST}}(\Vec{M},S)}{\sigma_{E,v,\mathrm{LST}}}\right)^2 \\- \sum_{\mathrm{LST},v}\ln{\left(2\pi\sigma_{E,v,\mathrm{LST}}^2\right)},
\end{multline}
where $T_{E,v,\mathrm{LST}}$ is the observed antenna temperature (absolute temperature measurment) at frequency $v$ and at a specific $\mathrm{LST}$, the term $\sigma_{E,v,\mathrm{LST}}$ is the reported uncertainty on this observation. $T_{\mathrm{mod},v,\mathrm{LST}}(\Vec{M},S)$ is the models' predicted antenna temperature (for a given $\Vec{M}$ and $S$), given by convolving the predicted sky with a beam model: 
\begin{equation}
    T_{\mathrm{mod},v,\mathrm{LST}}(\Vec{M},S) = \frac{1}{4\pi}\int_{0}^{4\pi}{B(\Omega,\phi_0,v)D_\mathrm{true}(\Omega,v,\mathrm{LST})\mathrm{d}\Omega}.
\end{equation}

The term $D_\mathrm{true}(\Omega,v,\mathrm{LST})$ is the model's predicted sky for frequency $v$ (rotated to the correct LST), given by equation \ref{eq:component_sep} for a specific $\Vec{M}$ and $S$. The term $B(\Omega,\psi_0,v)$ is the beam model at the correct frequency. We show the EDGES low band and high band beam models in figure \ref{fig:beams}. Note that before convolving we rotate the beam model to the observation latitude of $-26.7\degr$~\citep{Monsalve2021}, and to the azimuth angle used to take the observations; $\Psi_0=0\degr$ for the low-band~\citep{Mozdzen2018} and $\Psi_0=-5\degr$ for the high-band~\citep{Monsalve2021}.

Since we use chromaticity corrected spectral indexes, we must convolve with the beam models at the reference frequencies. Thus, for the $T$ vs LST curves between 40 and 100~MHz we convolve the predicted sky (at each of the frequencies) with the 75~MHz EDGES low-band beam model~\citep{Mahesh:2021rly}, and between 105 and 200~MHz we convolve with the 150~MHz EDGES high-band beam model (Mahesh et al. Private Communication).

Due to the high dimensionality of the joint posterior, it is impractical to draw samples directly. Instead, we marginalise over the distribution of component amplitudes, reducing the problem to sampling the marginal posterior of calibration and spectral parameters. 
\begin{equation}\label{marginalisation integral}
    P(E,D|a,\Vec{b},S) = \int{P(E,D|a,\Vec{b},\Vec{M},S)P(\Vec{M}|S)\mathrm{d}\Vec{M}}.              
\end{equation} 
Or equivalently;
\begin{multline}\label{marginalisation integral rewritten}
    P(a,\Vec{b},S|E,D) = \frac{P(D|a,\Vec{b},S)P(a,\Vec{b},S)}{P(E,D)} \\ \times \int P(E|\Vec{M},S) \times P(\Vec{M}|a,\Vec{b},S, D) \mathrm{d}\Vec{M}.
\end{multline}
As discussed in the first paper~\citep{Carter2025} for a set of diffuse maps each with $n_\mathrm{pix}$ pixels, the computational complexity of the analytical marginal likelihood found by evaluating the integral in equation~\ref{marginalisation integral} or equation~\ref{marginalisation integral rewritten} at best grows as $\mathcal{O}(n_\mathrm{pix}^{2.373})$~\citep{Davie2013}. To avoid this, we approximate the conditional distribution $P(\Vec{M}|a,\Vec{b},S, D)$ (which we can show to be a Gaussian with a analytically defined mean and covariance) as a delta function around its conditional mean set of component amplitude maps $\left<\Vec{M}|a,\Vec{b},S,D\right>$,
\begin{equation}
    P(\Vec{M}|a,\Vec{b},S, D) \approx \delta\left(\Vec{M} - \left<\Vec{M}|a,\Vec{b},S,D\right>\right),
\end{equation}
This approximation means that the marginal likelihood may be written as the product of $n_\mathrm{pix}$ individual pixel likehoods, resulting in a linear growth of computational complexity. The approximate marginal likelihood is then given as: 
\begin{multline}\label{approximatedion}
    P(a,\Vec{b},S|E,D) \approx\frac{P(D|a,\Vec{b},S)P(a,\Vec{b},S)}{P(E,D)} \\ \times P\left(E|\left<\Vec{M}|a,\Vec{b},S,D\right>,S\right).
\end{multline}
%\newpage
The term $P(D|a,\Vec{b},S)P(a,\Vec{b},S)$ is the analytically defined marginal likelihood of observing the diffuse dataset, $D$, for parameters $a,\Vec{b},S$. The term $P(E|\left<\Vec{M}|a,\Vec{b},S,D\right>,S)$ is EDGES likelihood (equation \ref{EDGES likelihood term}) evaluated for the conditional posterior mean set of component amplitudes, $\left<\Vec{M}|a,\Vec{b},S,D\right>$, and the spectral model, $S$.

We sample our approximate marginal distribution (equation~\ref{approximatedion}) using the PolyChord~\citep{polychord} implementation of the nested sampling algorithm~\citep{Skilling2004}, giving a set of marginal posterior samples, $\{a_i,\Vec{b}_i,S_i\}_\mathrm{posterior}$. For each of these marginal posterior samples, we then generate a sample set of component maps drawn from the conditional posterior distribution of the maps. This yields a set of posterior component map samples $\{\Vec{M}_i\}_\mathrm{posterior}$. Taken together, the marginal samples and the map samples, then form a set of samples drawn from the joint posterior distribution.

\section{Priors}
\begin{figure}
    \centering
    \includegraphics[width=0.99\linewidth]{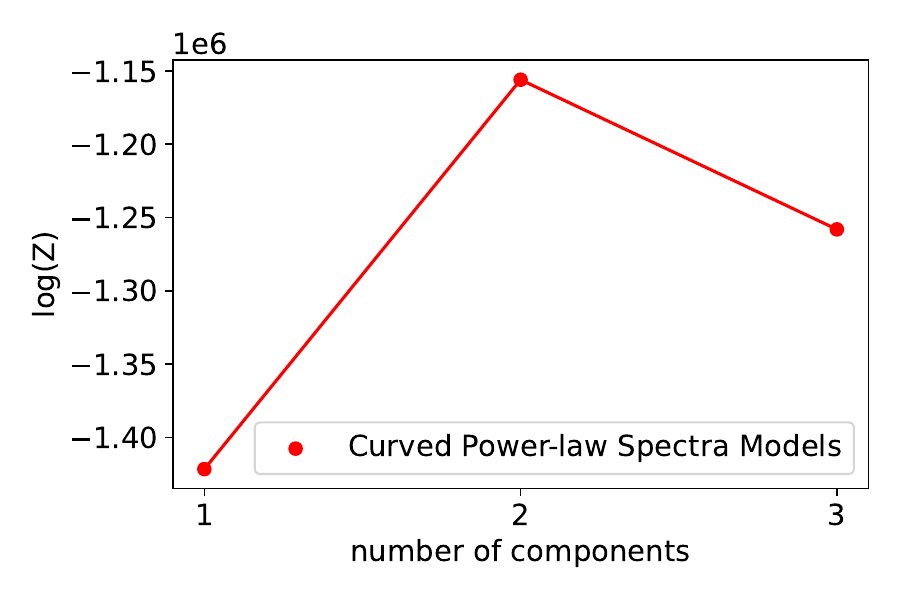}
    \caption{Bayesian evidence for a selection of different candidate models, computed via nested sampling of the marginal posterior using PolyChord~\citep{polychord}. We see that Bayesian evidence is highest for a two component model.}
    \label{fig:bayes-evidence}
\end{figure}

The prior for the component map amplitudes, $P(\Vec{M}|S)$, used for the marginalisation, is defined pixel-by-pixel as a Gaussian with mean of 0 and with covariance matrix $c_0(S)$:
\begin{equation}
    P(\Vec{M}_p|S)=N(\Vec{0},c_0(S)) \quad \forall \quad p \in \{1,\ldots,n_p\}.
\end{equation}
The map amplitude prior covariance matrix depends on the set of spectral parameters, $S$, and our prior assumptions about the sky covariance, it is defined as:
\begin{equation}
    c_0(S) = \left( S^T C_\mathrm{Sky}^{-1} S \right)^{-1}.
\end{equation}
The term $C_\mathrm{Sky}$ is our prior assumption for the covariance of the sky at the observed frequencies. We assume that the sky prior covariance matrix is diagonal and that the variance increases at lower frequencies according to a power-law. i.e. that the standard deviation of the sky prior at frequency $v$ is given by the power-law, $A_\mathrm{Sky}\left(v/v_\mathrm{Sky}\right)^{\beta_\mathrm{Sky}}$. The sky covariance matrix thus has diagonal elements given by:

\begin{equation}
    C_{\mathrm{Sky},{vv}} = A_\mathrm{Sky}^2\left(\frac{v}{v_\mathrm{Sky}}\right)^{2\beta_\mathrm{Sky}}.
\end{equation}
This choice of prior for the sky has two parameters, $A_\mathrm{Sky}$ the standard deviation (width) at the chosen reference frequency, and $\beta_\mathrm{Sky}$ the spectral index of the prior. We choose these parameters based on our prior assumptions about the sky temperature (at the reference frequency, $v_\mathrm{sky}$) and the sky's spectral behaviour. For the results shown in figure~\ref{fig:bayes-evidence}, we choose $ A_\mathrm{Sky}=400$ kelvin, $v_\mathrm{sky}=408~MHz$, and we assume a spectral index of $\beta_\mathrm{Sky}=-2.7$. This ensures that the sky prior remains broad across the frequency range 40-408~MHz. See the first paper~\citep{Carter2025} for a discussion of why we introduce the sky prior.

Note, unless otherwise indicated, all results presented in this paper use the following priors for the model parameters. For the spectral indexes of the component spectra, we use a uniform prior between -3.5 and +1, $P(\beta_c)=U(-3.5,1)$. For the spectral curvature of the component spectra, we use a Gaussian prior $P(\gamma_c)=N(0,1)$. For the zero level correction, we use a Gaussian prior of width 2000 kelvin at all frequencies $P(b_v)=N(0,2000~K)$. For the temperature scale correction, we use a uniform prior between 0.5 and 1.5 at all frequencies, $P(a_v)=U(0.5,1.5)$. Also note, for the component spectra, $S^c(v)$, we fix the reference frequency to be $v_0=120$~MHz for all models in this study (this reference frequency is distinct from the sky prior reference frequency).

\section{Bayesian Model Comparison}\label{s:Bayesian evidence}
We have a selection of possible models that could describe our data, $\{\mathcal{M}_i\}$. We use Bayesian model comparison~\citep{Trotta2008} to determine which of our candidate models is most likely given our observed dataset. Bayesian model comparison uses the denominator in Bayes theorem (known as the Bayesian evidence) to compare models by computing the posterior probability that a given model is true. In general for a dataset, $\mathrm{Data}$, and a specific model $\mathcal{M}_i$ with parameters $\theta_i$ applying Bayes theorem gives:
\begin{equation}
        P(\theta_i|\mathrm{Data},\mathcal{M}_i) = \frac{P(\mathrm{Data}|\theta_i,\mathcal{M}_i)P(\theta_i|\mathcal{M}_i)}{P(\mathrm{Data}|\mathcal{M}_i)} 
\end{equation}
If we perform nested samping for a particular model, $\mathcal{M}_i$, we obtain both samples from the posterior distribution of the model parameters, $\theta_i$, and an estimate of the Bayesian evidence for that model, $P(\mathrm{Data}|\mathcal{M}_i)$~\citep{Skilling2004}. Thus, if we have performed nested sampling for each model in our set \(\{\mathcal{M}_i\}\), we can determine the probability that model \(\mathcal{M}_i\) is correct given the observed data by applying Bayes’ theorem again at the model level:  
\begin{equation}
    P(\mathcal{M}_i|\mathrm{Data}) = \frac{P(\mathrm{Data}|\mathcal{M}_i)P(\mathcal{M}_i)}{P(\mathrm{Data})} = \frac{P(\mathrm{Data}|\mathcal{M}_i)P(\mathcal{M}_i)}{\sum_i P(\mathrm{Data}|\mathcal{M}_i)},
\end{equation}
\subsection{Results for Bayesian Model Comparison}
For B-GSM each of our candidate models, has a different number of emission components and potentially a different parametrisation for the component spectra. In this study we investigated models with 1, 2, and 3 emission components. We choose to restrict ourselves to only using models with curved power-law component spectra;
\begin{equation}\label{curved powerlaw}
    S^c(v) = \left(\frac{v}{v_0}\right)^{\beta_c + \gamma_c\log\left(v/v_0\right)}.
\end{equation}
Investigation of alternative parametrisations for the spectral model will be left for future work. We assume that each of these candidate models is a-priori equally likely. 

For each candidate model we performed nested sampling of the approximate marginal posterior (equation \ref{approximatedion}), giving both a set of posterior samples and a Bayesian evidence value for each of the candidate models. Note that the both the diffuse dataset and the EDGES datasets along with all priors are kept identical between the candidate models, and nested sampling used $n_\mathrm{live}=500$ and $n_\mathrm{repeat}$ equal to 5 times the number of model parameters. In figure~\ref{fig:bayes-evidence} we plot the Bayesian evidence values for each of the three candidate models, we see that a two component model is strongly favoured.

\section{Posterior For Highest Evidence Model}\label{s:highest evidence posterior}

\begin{figure*}
    \centering
    \includegraphics[width=0.86\linewidth]{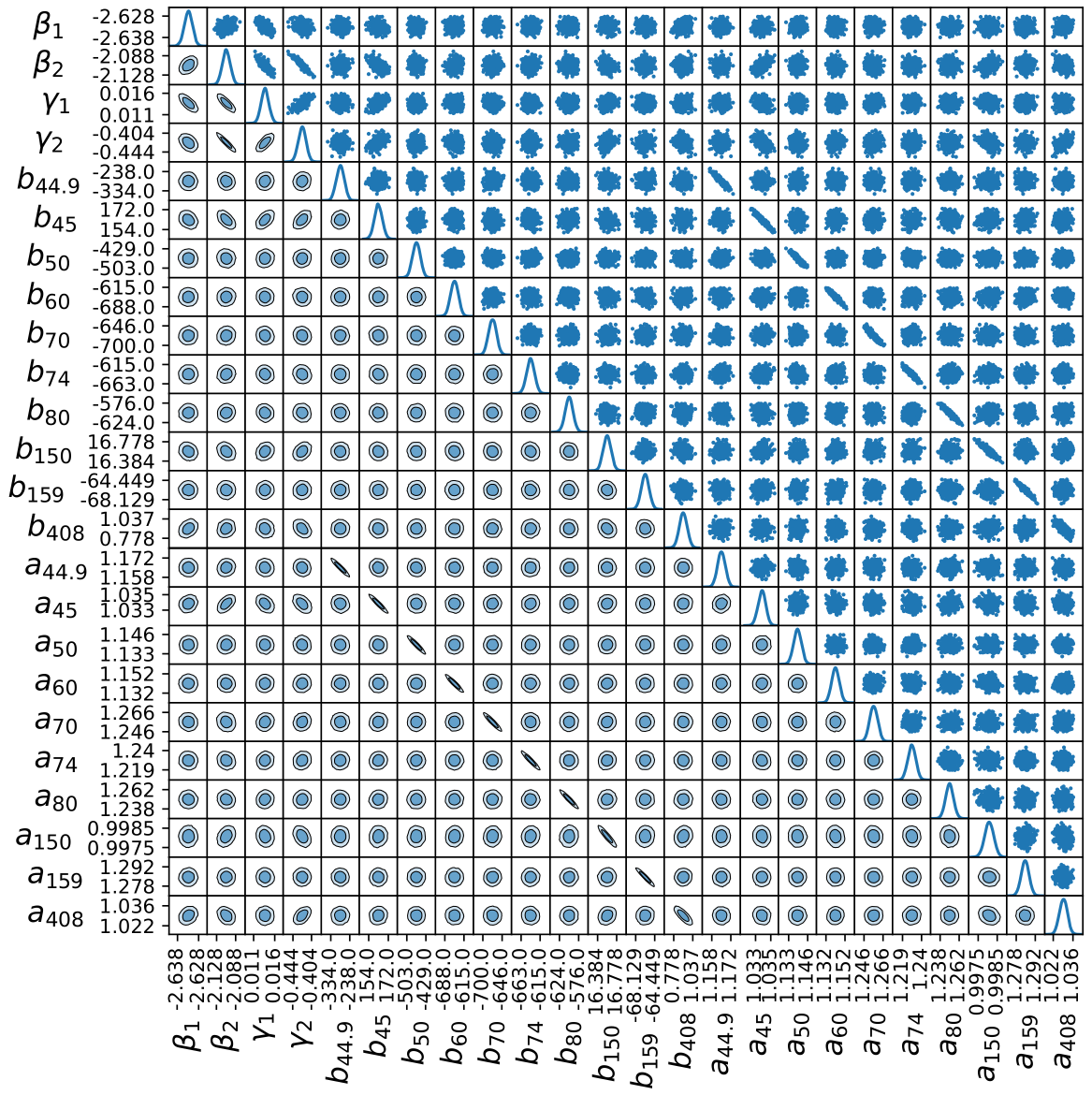}
    \caption{Marginal posterior of spectral and calibration parameters for the highest evidence model, figure produced using \texttt{Anesthetic} \citep{anesthetic}. The first two parameters $\beta_1$, $\beta_2$, are the spectral index for each of the two component spectra. The next two parameters $\gamma_1$, $\gamma_2$ are the spectral curvature for each of the two component spectra. The final 20 parameters are the corrections to the temperature zero levels $b_v$ and temperature scales $a_v$ for each of the 10 diffuse maps in the diffuse dataset.}
    \label{fig:marginal_posterior}
\end{figure*}

Figure \ref{fig:marginal_posterior} shows the set of samples, $\{a,\Vec{b},S\}_\mathrm{posterior}$, drawn from the marginal posterior, of our highest evidence model. The first four parameters correspond to the spectral model; $\beta_1$ and $\beta_2$, are the spectral indexes, and $\gamma_1$ and $\gamma_2$, are the spectral curvature for each of the two components. The next ten parameters are the zero-level corrections, $b_v$, for each of the diffuse maps. The final ten parameters are the temperature-scale corrections, $a_v$, for each of the diffuse maps. We show the posterior mean and standard deviation, for each parameter, in table~\ref{t:mean_post}. 

\begin{table*}
    \centering
    \caption{Posterior mean and standard deviation (summary statistics for the marginal posterior in figure~\ref{fig:marginal_posterior})}
    \begin{tabular}{cc|cc}
        Parameter & Posterior & Parameter & Posterior\\
        \hline\\
        $\beta_1$ &  -2.633 $\pm$ 0.002 & $b_{159}$ & -66.3 K $\pm$ 0.7 K\\
        $\beta_2$ &  -2.108 $\pm$ 0.008 & $b_{408}$ & 0.91 K $\pm$0.05 K\\
        $\gamma_1$ & 0.014 $\pm$ 0.001 & $a_{44.9}$ & 1.165 $\pm$0.003\\
        $\gamma_2$ & -0.424 $\pm$ 0.008 & $a_{45}$ & 1.0338 $\pm$0.0005\\
        $b_{44.9}$ & -286 K $\pm$ 19 K & $a_{50}$ & 1.139 $\pm$0.003\\
        $b_{45}$ & 162 K $\pm$ 3 K & $a_{60}$ & 1.142 $\pm$0.004\\
        $b_{50}$ & -466 K $\pm$ 15 K & $a_{70}$ & 1.256 $\pm$ 0.004\\
        $b_{60}$  & -651 K $\pm$ 15 K & $a_{74}$ & 1.229 $\pm$ 0.004\\
        $b_{70}$ & -673 K $\pm$ 11 K & $a_{80}$ & 1.250 $\pm$0.005\\
        $b_{74}$ & -639 K $\pm$ 10 K &$a_{150}$ & 0.9981 $\pm$ 0.0002\\
        $b_{80}$ & -600 K $\pm$ 10 K & $a_{159}$ & 1.285 $\pm$0.003\\
        $b_{150}$ & 16.58 K$\pm$ 0.08 K&$a_{408}$ & 1.029 $\pm$ 0.003\\
        
    \end{tabular}
    \label{t:mean_post}
\end{table*}

\begin{figure*}
    \includegraphics[width=0.99\linewidth]{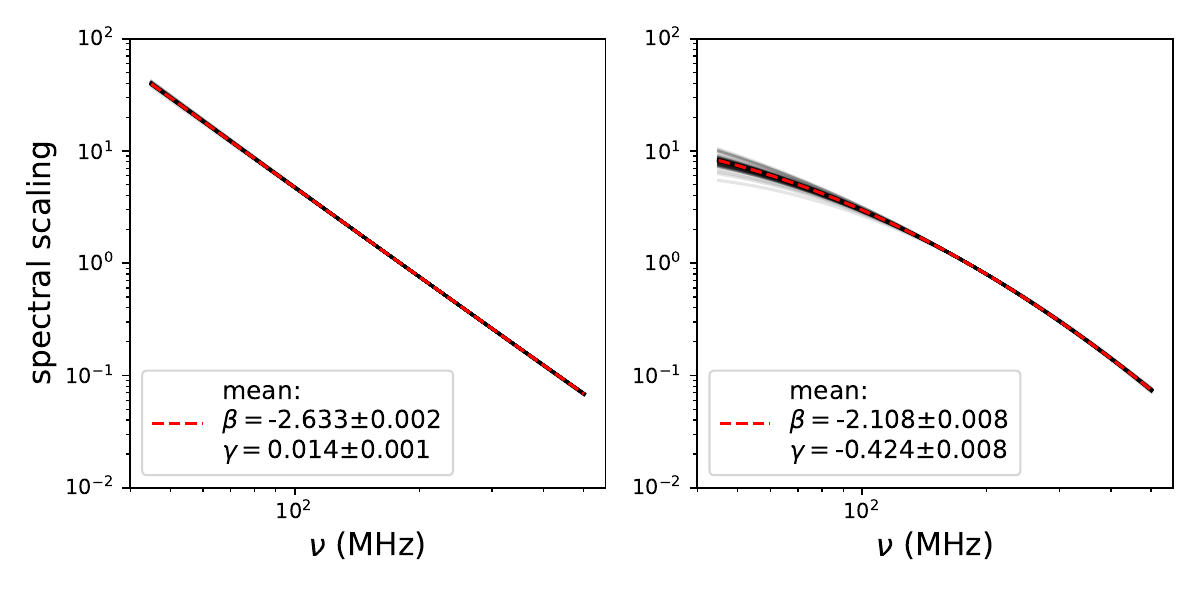}
    \caption{Functional posterior plots for the two component spectra, figure plotted with fgivenx \citep{fgivenx}. We see that the posterior for the  second component spectra is broader than the first component, indicating a greater uncertainty. The first component follows a power-law spectrum with spectral index $\beta_1=-2.633\pm0.002$ with curvature $\gamma_1=0.014\pm0.001$, approximately corresponding with previous reports of the synchrotron spectral index~\citep{Spinelli2021,Guzman2010,Lawson1987}. The second component follows a power-law with spectral index $\beta_2=-2.108\pm0.008$ and curvature $\gamma_2=-0.424\pm0.008$.}
    \label{fig:posterior_specs}
\end{figure*}

We find that the first component spectrum has a spectral index of $\beta_1=-2.633\pm0.002$ and a curvature of $\gamma_1=0.014\pm0.001$, the spectrum for the second component is found to have $\beta_2=-2.108\pm0.008$ and $\gamma_2=-0.424\pm0.008$. Note that the spectral parameters are correlated: \(\gamma_1\) and \(\gamma_2\) are positively correlated, \(\gamma_1\) is negatively correlated with both \(\beta_1\) and \(\beta_2\), while \(\gamma_2\) is negatively correlated with \(\beta_2\) but not \(\beta_1\). The functional forms of the posterior spectra, for the two emission components, are plotted in figure~\ref{fig:posterior_specs}. Each black line in the figure is the spectrum produced for a specific posterior sample, the red dashed lines show spectra for the posterior mean set of spectral parameters. 

In the case of the calibration corrections, we see that for each frequency the correction to the temperature scale, $a_v$, and the correction to the temperature zero-level, $b_v$, are negatively correlated. No correlation is apparent between calibration parameters at different frequencies. This behaviour is expected and was also seen when validating our approach on synthetic data~\citep{Carter2025}. 

The posterior calibration corrections for the Guzman 45~MHz map are \( b_{45} = +162 \pm 3 \) K for the zero-level correction and \( a_{45} = 1.0338 \pm 0.0005 \) for the temperature scale correction. For the LW 150~MHz map, we find \( b_{150} = +16.58 \pm 0.008 \) K and \( a_{150} = 0.9981 \pm 0.0002 \).  

Ideally, given the calibration corrections that we applied during pre-processing of the maps (which are derived from EDGES data \cite{Monsalve2021}), we would expect to recover a zero-level correction of 0~K and a scale correction of 1 for both maps. While the scale corrections we determine are close to 1, the zero-level corrections deviate significantly from the expected value. These discrepancies are likely due to the fact that in this study we do not have access to raw EDGES data, whereas the corrections used in the pre-processing are determined using raw EDGES data~\citep{Monsalve2021}. 

It is notable that, the corrections for the 150~MHz map are closer to the expected result than those for the 45~MHz map. This can be explained by the fact that 150~MHz is the reference frequency for the EDGES high-band spectral indexes, as such we have an actual EDGES $T$ vs LST curve for this frequency (taken from \cite{Monsalve2021}). At 45~MHz, we relied on the 75~MHz curve rescaled using EDGES low-band spectral indices. It is therefore unsurprising that our posterior calibration corrections at 150~MHz are closer to the expected result, compared to the 45~MHz corrections. We should also note that, as we do not have access to raw EDGES data, we had to digitise the published graphs of both EDGES $T$ vs LST curves \citep{Monsalve2021, Mozdzen2018} and spectral indexes~\citep{Mozdzen2016,Mozdzen2018}. Errors in the antenna temperatures and spectral indexes introduced by this digitisation are likely to have contributed to the mismatch in the posterior and expected calibration corrections at 45~MHz and 150~MHz. Unfortunately, without access to the original EDGES data, the results presented here represent the best achievable with publicly available data.

For the LWA1 maps at 44.9 50, 60, 70, 74, and 80~MHz we see that the posterior mean temperature scale corrections are of order $\sim16-25\%$ and the zero level corrections are found to be of order a few hundred kelvin. These corrections, to the LWA1 maps, approximately align with the reported $\sim15\%$ disagreement between LWA1 maps and absolute temperature measurements taken using the LEDA instrument~\citep{Spinelli2021}. Additionally, for the EDA2 159~MHz map, we find posterior calibration corrections of $a_{159}=1.285\pm0.003$ for the temperature-scale and $b_{159}=-66.3\pm0.7$ K for the zero-level. 
%\newpage

Finally, for the Haslam 408~MHz map we determine posterior calibration corrections of $a_{408}=1.029\pm0.003$ for the temperature-scale and $b_{408}=+0.91\pm0.05$ K for the zero-level. This agrees well with the widely reported uncertainties on the temperature-scale and zero-level of $\leq10\%$ and $\pm3$ K respectively~\citep{Remazeilles2015}. However, our posterior calibrations for Haslam disagree strongly with the 60\% gain correction reported in \cite{HaslamCalibration2024}. In this study, we have assumed that the temperature-scale calibration can be corrected using a single global correction factor for each map. In contrast, \cite{HaslamCalibration2024} use Bayesian model comparison to test for multiplicative gain biases across three sky regions, which allows for variations in the calibration errors as a function of position on the sky. This methodological difference may explain the strong disagreement between our posterior temperature-scale correction for Haslam of $\sim 3\%$ and that reported by \cite{HaslamCalibration2024}. 

\begin{figure*}
    \centering
    \includegraphics[width=0.8\linewidth]{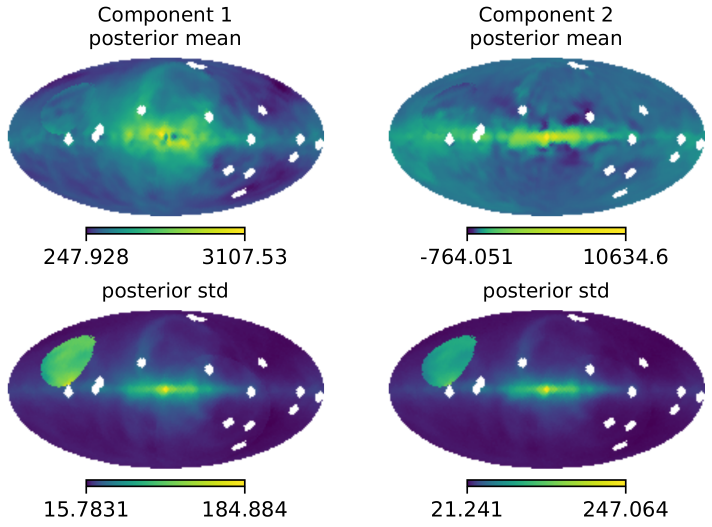}
    \caption{Top row shows the (weighted) posterior mean component amplitude maps for each of the two components. Bottom row shows the (weighted) posterior standard deviation (statistical uncertainty) for the component amplitudes. The first component appears to correspond to Galactic synchrotron emission.}
    \label{fig:posterior_comps}
\end{figure*}

\subsection{Posterior Emission Components}
So far, we have only discussed the marginal posterior of the model parameters.
We will now discuss the posterior distribution of the component amplitude maps.
As mentioned in section~\ref{s:brief theory}, to generate a set of posterior samples of the component amplitudes, $\{\Vec{M}_i\}_\mathrm{posterior}$, we use our set of samples from the marginal posterior, $\{a_i,\Vec{b}_i,S_i\}_\mathrm{posterior}$. Specifically, we generate a sample set of component amplitude maps for each of the samples in the marginal posterior. The $i$-th posterior sample amplitude maps, $M_i$, is drawn from the conditional distribution, $M_i \sim P(M|a_i,\Vec{b}_i,S_i, D)$, conditioned on the $i$-th sample from the marginal posterior $a_i,\Vec{b}_i,S_i$ and our observed data. This gives us a set of sample component amplitudes drawn from the posterior distribution 

In figure \ref{fig:posterior_comps} we show the weighted mean and standard deviation for these component amplitude posterior samples. The first row shows the weighted mean amplitude map for each of the two components, and the second shows the $1\sigma$ uncertainty (posterior standard deviation) for each component. We must note that these component amplitudes are for a reference frequency of $v_0=120$~MHz and are shown in units of kelvin. The maps are shown in a Mollweide projection, the temperature scale is linear for temperatures below 300 kelvin and logarithmic for temperatures above 300 kelvin (used in order to display the negative temperatures for the second component). 

The second component displays non-physical negative temperatures in certain regions of the sky. A possible explanation for these negative temperatures lies in the fact that B-GSMs component separation implicitly assumes that spectral behaviour for a component is identical for all regions of the sky. This is not a physically motivated or realistic assumption, previous studies have shown variation in the Galactic spectral behaviour as a function of position~\citep{Guzman2010,Lawson1987}. It is possible that these negative temperatures in the second component are simply accounting for an unmodeled spatial variation in the spectrum of the first component.

We note that each of the two components is dominant for a different region of the galaxy. The first component has large amplitudes up to high Galactic latitudes, particularly in the northern polar spur. The second component is dominant in the galactic plane, with far smaller contributions at the high Galactic latitudes. This is unsurprising, the second component has a flatter spectrum than the first, and previous studies~\citep{Guzman2010,WMAPsynchrotronSpectrum} have shown that the spectral index of Galactic emission is flatter (less negative) in the Galactic plane. This further supports the idea that the second component is accounting for spatial variation in spectral behaviour of the first.

The posterior standard deviation for both component amplitude maps is largest in the Galactic plane, Galactic centre, and the northern sky. The large uncertainty in the Galactic plane and centre is unsurprising, as these are by far the brightest regions of the sky. However, the large uncertainty in the northern sky is a surprising result, given that there are nine diffuse maps covering the full northern sky, compared to just four covering the full southern sky. Intuitively, having more independent datasets should lead to greater constraint on the posterior and lower uncertainty. 

A possible explanation, for the large northern sky uncertainty, lies in the fact that our posterior is conditioned on both the diffuse maps and the EDGES observations. The EDGES observations only cover the southern sky, and are used to inform our inference of the spectral behaviour and the calibration corrections for the diffuse maps. As such the posterior spectral parameters and posterior calibration corrections are primarily determined by the southern sky observations.

Additionally, we made the assumption that the calibration corrections for each diffuse map (the zero-level offset $b_v$ and temperature scale correction $a_v$) are uniform across the entire sky. This assumption is likely incorrect, both the Haslam 408 MHz and LW 150 MHz maps were constructed by combining multiple surveys, each covering different regions of the sky. It is unlikely that a single global calibration correction can accurately account for variations in the calibration across all regions of these maps.

Since our calibration parameters are constrained using southern sky data, they likely do not fully correct for calibration errors in the northern sky. As a result, when we sample the posterior component amplitudes using the conditional distribution $P(\Vec{M} | a,\Vec{b},S,D)$, the parameters we condition on $(a,\Vec{b},S)$ are biased toward southern sky observations. This could potentially lead to a wider posterior distribution (more uncertainty) for the component amplitudes in the northern sky. 

This issue could potentially be addressed by conditioning our posterior on an additional set of absolute temperature measurements from an instrument located in the northern hemisphere. For example, we could condition the posterior on both the EDGES dataset (used in this study) and on LEDA observations of the sky's spectral behaviour~\citep{Spinelli2021}. Investigation of this was beyond the scope of this study and will be left for future work.

We note that both components display a non-physical hard boundry in the component mean (this boundry is at the edge of the poorly observed region of the northern sky). This is an artefact of plotting a point estimate (the mean), and would be masked by plotting samples including noise.

\subsubsection{Physical interpretation of spectra and components}
As we have previously discussed, the spectral index of the sky varies as a function of position, with the spectrum becoming flatter in the galactic plane~\citep{Guzman2010,WMAPsynchrotronSpectrum}. Given that B-GSM restricts each of its components to have the same spectral behaviour for all regions of the sky, it is unlikely that the posterior emission components correspond directly to physical Galactic emission components. Despite this we will briefly discuss possible physical interpretations of the components identified by B-GSM.

The low-frequency sky is expected to be dominated by Galactic synchrotron emission~\citep{Lian2020}, which is expected to follow a power-law spectrum. The synchrotron spectral index is reported to be between $-2.6<\beta<-2.5$ for frequencies in the range 45~MHz and 408~MHz~\citep{Guzman2010}, and a slightly flatter spectral index of $\beta=-2.5\pm0.1$ is reported between 50~MHz and 87~MHz~\citep{Spinelli2021}. We see that
the first component's spectrum is broadly in agreement with these literature values for the synchrotron spectral index. Additionally, looking at the posterior amplitude map for the first component (left panels of figure~\ref{fig:posterior_comps}), we can clearly see the northern polar spur, which is known to be dominated by synchrotron emission~\citep{GSM}. It seems probable that the first component approximately represents Galactic synchrotron emission. 

The posterior amplitude map for the second component (right panels of figure~\ref{fig:posterior_comps}) displays non-physical negative temperatures, and most likely does not directly correspond to any individual physical emission component. As previously discussed, these negative amplitudes in regions above and below the Galactic plane in the second component are likely accounting for spatial variation in the spectral behaviour of the first (Galactic synchrotron) component. Within the Galactic plane we expect to see some contribution from Galactic free-free emission~\citep{Lian2020}, this is likely also captured within the second component. As such, we interpret the second component as modelling spatial variation in the synchrotron spectral index, and also containing a contribution from free-free emission.

\begin{figure*}
    \centering
    \includegraphics[width=0.77\linewidth]{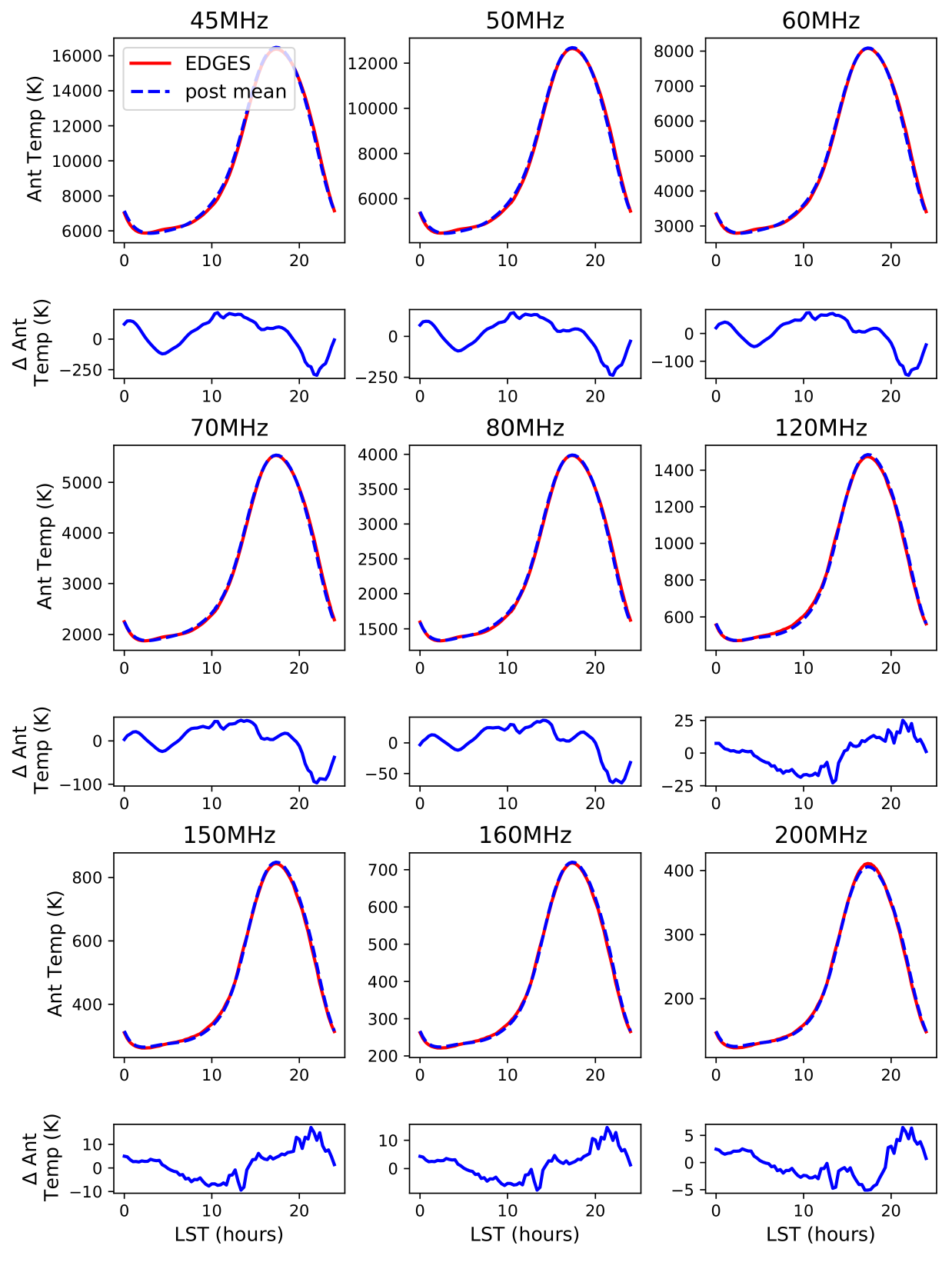}
    \caption{Comparison of posterior mean $T$ vs LST curves and the $T$ vs LST curves from EDGES. We see that across the frequency range 45-200~MHz the posterior temperatures are in excellent agreement with the EDGES data. Indicating that B-GSMs posterior is correctly calibrated across this frequency range.}\label{fig:posterior of TvsLST}
\end{figure*}

\subsection{Posterior Sky Predictions}
Given our spectral and component amplitude posterior samples, we can now generate samples from the posterior distribution of the sky. For a given frequency $v$, we construct the $i$'th posterior sky sample, $\mathrm{Sky}_{i,v}$, using the corresponding $i$'th posterior sample of spectral parameters, $S_i$, and component amplitudes, $M_i$. This process is repeated for each posterior sample $ S_i \in \{S_i\}_\mathrm{posterior}$ and $M_i \in \{M_i\}_\mathrm{posterior}$, producing a set of posterior sky samples $\{\mathrm{Sky}_{i,v}\}_\mathrm{posterior}$ , with each sample generated according to equation~\ref{eq:component_sep}. This allows us to produce a full posterior distribution of sky predictions at any frequency within the range 45-408 MHz.

To assess the calibration of the posterior sky predictions, we generate a full sky posterior at 45, 50, 60, 70, 80, 120, 150, 160, and 200~MHz. Each of these posterior sky predictions is then convolved with the EDGES beam model (for LSTs covering the full range 0 to 24 hours) in order to produce a posterior set of predicted $T$ vs LST curves for B-GSM at each of these frequencies. In figure~\ref{fig:posterior of TvsLST}, we plot the mean posterior predicted $T$ vs LST curve at each frequency and compare it with a $T$ vs LST curve (at the same frequency) from our EDGES dataset. The larger panels show the $T$ vs LST curves (in units of kelvin) with the EDGES dataset shown in red and the B-GSM predictions shown in blue. The smaller panels show the residuals between the predictions and EDGES. 

We see that for all tested frequencies, the posterior predicted and EDGES $T$ vs LST curves are in excellent agreement across the full 24 hours of LST. The residual between B-GSMs predicted antenna temperatures and the EDGES dataset is very small (the largest percentage difference is $<3.2\%$). This indicates that B-GSMs posterior sky predictions have been successfully calibrated to the EDGES dataset.

\begin{figure}
\begin{center}
    \includegraphics[width=0.9\linewidth]{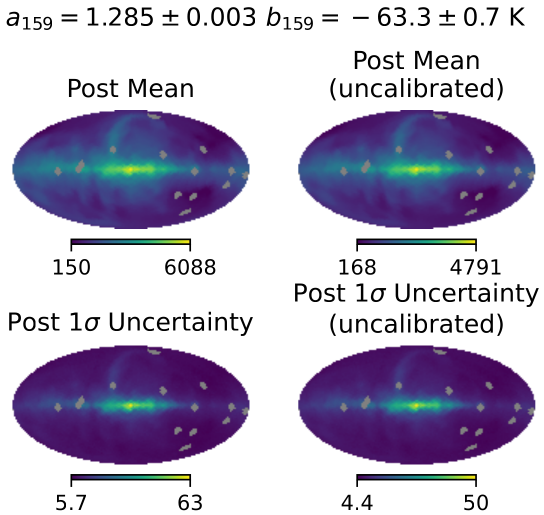}
\end{center}
\caption{Comparison of the original posterior sky prediction at 159~MHz, and the posterior sky prediction after removing the calibration corrections for this frequency. The posterior after removing calibration correction (the ``uncalibrated" posterior) is used to compare model predictions with the observed diffuse sky maps in our dataset. All maps are shown in Galactic coordinates in units of kelvin on a log scale.\label{fig:calibration removed}}
\end{figure}
%\newpage
The posterior sky samples, $\{\mathrm{Sky}_{i,v}\}_\mathrm{posterior}$, describe B-GSM's prediction for the sky for the frequency $v$. However, it is difficult to make a direct comparison between this posterior sky prediction and the diffuse observations we used to construct our model. This is because we can only make comparisons to the observations of the sky contained in our diffuse dataset. These observations have uncertainties due to thermal noise and calibration. In order to make a comparison between our predicted sky and the observed sky, we must recall the basis of our model:

\begin{align}
    &a_v D_v(\Omega) + b_v = \left(\sum_{c=1}^{k}{M_c(\Omega)S^c(v)}\right)  + a_v N(\Omega,v)\\
    &\implies N(\Omega,v) = D_v(\Omega) - \frac{\left(\sum_{c=1}^{k}{M_c(\Omega)S^c(v)}\right)-b_v}{a_v} ,\label{decalibrated eq}
\end{align}
i.e. the sky after applying calibration is equal to our model prediction plus the (calibrated) noise in the sky observations. Rearranging, we see that the noise in the observed sky at frequency $v$, $N(\Omega,v)$, can be written as the residual between the observed sky, $D_v(\Omega)$, and the predicted sky after removing our calibration corrections (equation \ref{decalibrated eq}).

This ``uncalibrated" posterior sky is defined by the samples, $\mathrm{Sky}_{\mathrm{uncal},i,v} = (1/a_{v,i})(S_i M_i - b_{v,i})$. Where (for frequency $v$) the $i$'th ``uncalibrated" posterior sky sample is computed using the inverse of the $i$'th posterior scale correction $a_{v,i}$, the $i$'th posterior zero correction $b_{v,i}$, and the $i$'th set of spectral parameters and component amplitudes. This gives us a set of sample sky predictions from the ``uncalibrated'' posterior. In figure \ref{fig:calibration removed} we show both the original posterior sky prediction (mean and standard deviation) and the ``uncalibrated" posterior sky prediction (mean and standard deviation) for B-GSM at 159~MHz.

\begin{figure*}
    \centering
    \includegraphics[width=0.78\linewidth]{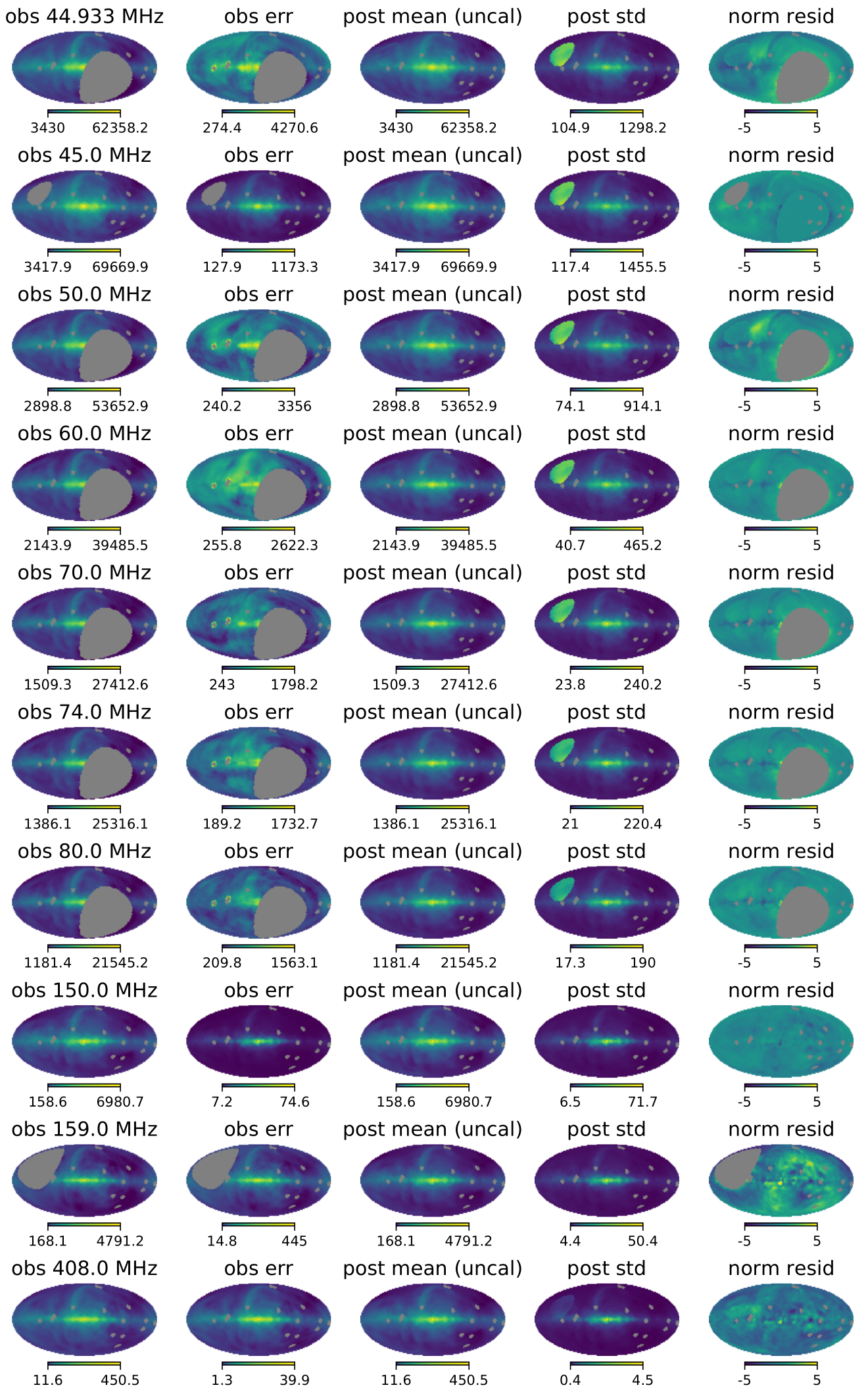}
    \caption{Posterior sky predictions and comparison with the observed sky for frequencies between 45 and 408 MHz. At all frequencies, the posterior mean is visually indistinguishable from the observed sky, though the normalised residuals show structure.}\label{fig:post sky}
\end{figure*}

We generate samples of the ``uncalibrated'' posterior predicted sky at each of the frequencies observed in our diffuse dataset. In figure~\ref{fig:post sky} we compare the observed sky and the mean ``uncalibrated'' posterior predicted sky at each frequency. The first column shows the observations, the second the reported (or assumed) uncertainty maps, third is the ``uncalibrated'' posterior prediction, and fourth is the posterior standard deviation (uncertainty on the prediction). We see that the posterior predictions and the observations are indistinguishable by eye, with the same spatial structure and brightness temperatures. 

Looking at the posterior uncertainty maps, we see that the region of greatest uncertainty changes for the different frequencies. At the lowest frequencies, the uncertainty is largest in the northern sky. As previously discussed, a possible explanation for this large northern sky uncertainty is the fact that the EDGES dataset (against which we calibrate) only has southern sky coverage. Potentially resulting in calibration corrections that do not properly account for errors in the northern sky.

In the fifth column of figure~\ref{fig:post sky} we show the normalised residuals between the predicted and observed sky. These are defined as $\left(\left<\mathrm{Sky}_{\mathrm{uncal,i,v}}\right>-D_v(\Omega)\right)/{N(\Omega,v)}$, with the average being over the set of posterior samples (this expression is found by rearranging equation~\ref{decalibrated eq}). Note that the residuals are normalised using the noise on the observed data, not the $1\sigma$ uncertainty on the posterior prediction. 

Looking at the spatial distribution of the normalised residuals (column 5) we see there is spatial structure to their distributions, they are not white noise. The mean posterior sky is over-predicting the temperature in the galactic plane at 45~MHz and is under-predicting the temperature in the galactic plane for all of the LWA1 maps (44.9, 50, 60, 70, 74, 80~MHz). Additionally, the temperature of the southern sky is over-predicted relative to the EDA2 map at 159~MHz.

\begin{figure}
    \centering
    \includegraphics[width=\linewidth]{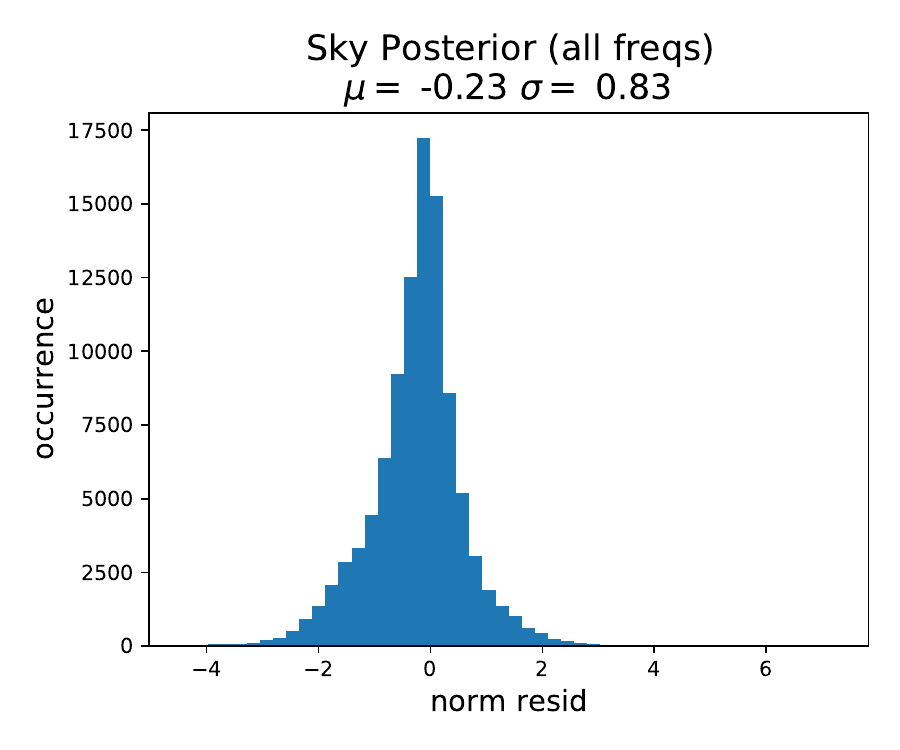}
    \caption{Histogram of the normalised residuals between the observed sky and the mean posterior predicted sky. The normalised residuals shown are the combination of all the frequencies in figure \ref{fig:post sky}. The normalised residuals do not follow a Gaussian distribution, and have a mean value of $\mu=-0.23$ and standard deviation of $\sigma=0.83$.\label{fig: sky post histo}}
    \label{fig:enter-label}
\end{figure}

Figure \ref{fig: sky post histo} shows a histogram of the combined set of normalised residuals for all tested frequencies. The distribution of normalised residuals is only roughly Gaussian, and has a mean of -0.23 and a standard deviation of 0.83. The non-zero mean suggests that B-GSM is systematically under-predicting sky temperatures. The standard deviation being smaller than 1 may suggest that the estimated uncertainty for the observed diffuse maps is too large.

A possible explanation for the non-Gaussian distribution of the normalised residuals could lie in the assumptions we made about the properties of the noise in the observed maps. Our likelihood function, assumes that the additive noise in the observed diffuse maps is Gaussian distributed. This should be approximately true for the thermal noise, however the uncertainty maps we used for the 45~MHz and 150~MHz maps come from the calibration uncertainty not the thermal noise~\citep{Monsalve2021}. Additionally, we assumed that noise between pixels is uncorrelated. This assumption is definitely incorrect for both the thermal noise and the calibration uncertainties. Thermal noise will be correlated on a scale equal to the FWHM of the observing beam, and calibration errors are necessarily correlated across the whole sky. It is possible that our failure to account for the correlations in the noise is the cause of the non-Gaussian distribution for the normalised residuals.

\section{Conclusions}\label{s:conclusions}
In this study, we have presented the results for development of a new calibrated low-frequency foreground model, the Bayesian Global Sky Model (B-GSM). We have employed a joint Bayesian analysis of both diffuse emission surveys and EDGES absolute temperatures, to perform simultaneous calibration and component separation. Our approach allows for rigorous quantification of uncertainties in foreground modelling, and ensures absolute calibration for the predicted sky.

Using Bayesian model comparison, we determine that the low-frequency sky is optimally modelled by two emission components each following a power-law spectrum. The results are consistent with a foreground that is dominated by Galactic synchrotron. The first component's spectrum is consistent with literature values for the synchrotron spectral index~\citep{Spinelli2021, Guzman2010}. The second component appears to model spatial variation in the synchrotron spectrum and may contain a contribution from free-free emission. 

We find that the Haslam 408~MHz map is well calibrated, requiring a temperature scale adjustment of $1.029\pm0.003$ (approximately 3\%) and an adjustment to its zero-level of $0.91\pm0.05$ kelvin. These calibration corrections are consistent with the reported uncertainties for Haslam \citep{Remazeilles2015}, but strongly disagree with the 60\% gain correction reported by \cite{HaslamCalibration2024}. We attribute this discrepancy to differences in the methodology, our approach fits for a single global calibration correction for each of the observed maps, whereas \cite{HaslamCalibration2024} accounts for spatially varying calibration error.

The posterior predicted absolute temperatures for the sky are in excellent agreement with the EDGES dataset. We find that the posterior $T$ vs LST curves have the same shape and amplitude as EDGES across the frequency range 45 MHz to 200 MHz. The posterior predicted $T$ vs LST curves agree with EDGES at a <3.2\% level, for all LSTs and frequencies in this range. This demonstrates that B-GSMs achieves accurate absolute calibration for its posterior sky predictions. 

The posterior predicted spatially resolved diffuse sky is indistinguishable by eye from the observed sky maps. However, we find that the normalised residuals (between the posterior predicted sky and observations) show structure and are not Gaussian distributed. The structure in the normalised residuals may be due to spatial variations in the spectral behaviour that are not fully accounted for. The non-Gaussian distribution of the normalised residuals may be due to correlations in the noise between pixels that are not modelled in B-GSM.

We should note that the uncertainty in posterior predictions for the northern sky is larger than expected. We attribute this to our assumption that calibration corrections are uniform across each diffuse map, and to the fact that the EDGES dataset only covers the southern sky. This results in calibration corrections that do not account for spatial variation in the calibration error for the maps in the diffuse dataset.

Overall, despite its limitations, B-GSM is able to provide full posterior predictions for the diffuse sky at any frequency between 45 and 408~MHz, ensuring robust uncertainty quantification. By conditioning on both diffuse and absolute temperature dataset, it overcomes key issues present in previous sky models and achieves absolute temperature calibration at a $\leq3.2\%$ level relative to the absolute temperature dataset (EDGES).

\section{Further Work}
Future research on B-GSM will focus on addressing several limitations identified in this study. In particular we will aim to address remaining issues with calibration, reducing uncertainty in posterior predictions, and improving our modelling of instrumental noise. 

A future goal is to address remaining issues with the calibration of the diffuse emission dataset used in B-GSM. In this work we assumed that the calibration correction at each frequency, can be achieved using a single global temperature scale and zero-level correction $a_v$ and $b_v$, that are applied uniformly across the entire sky. However, as previously discussed several maps in the diffuse dataset (e.g. the Haslam 408~MHz and LW 150~MHz maps) are composites of multiple surveys each covering different regions of the sky. As such, it is likely that different regions of the sky will require different corrections to their temperature scale and zero-levels. We could account for this by allowing spatial variation in the calibration corrections for each frequency. One option is that for the observed map at each frequency we could fit for two pairs of calibration corrections $a_{v,\mathrm{north}}$, $b_{v,\mathrm{north}}$ and $a_{v,\mathrm{south}}$, $b_{v,\mathrm{south}}$. A better option would be to obtain the original partial sky coverage surveys (which where each taken on a single instrument) and fit global calibration corrections for each of these partial sky surveys.

Accounting for spatial variation in the calibration corrections would require us to condition the posterior on two absolute temperature datasets covering both the northern sky and the southern sky. We could do this by including both EDGES absolute temperature data and LEDA~\citep{Spinelli2021} data. This would allow inference of the calibration corrections that are informed by data from both hemispheres. This future version of B-GSM would therefore be conditioned on three independent datasets; EDGES southern sky absolute temperature data, LEDA northern sky absolute temperature data, and our existing dataset of ten diffuse emission surveys. The additional data would provide an independent set of constraints on the calibration parameters and spectral behaviour, specifically focused on the northern sky. This has the potential to lead to reduced uncertainty in the northern sky posterior predictions, and improve calibration accuracy. 

Additionally, we aim to address calibration issues introduced by our use of an approximate EDGES dataset. In this study we did not have access to the original EDGES $T$ vs LST measurements, and instead had to rescale $T$ vs LST measurements taken at a reference frequency using spectral indexes. Both the spectral indexes and the reference frequency $T$ vs LST measurements, used in this study, were digitised from the published graphs in \citep{Monsalve2021,Mozdzen2016,Mozdzen2018}. This digitisation introduced errors into the absolute temperature dataset, and as discussed earlier will have resulted in slight mis-calibration of B-GSMs posterior. We see this in the posterior calibration corrections found for the Guzman 45 MHz and LW 150 MHz maps which do not agree within uncertainty with the expected 0 K zero level correction and scale correction of 1. To address this issue we would ideally aim to repeat our analysis using the original EDGES $T$ vs LST measurements, if we are given access to this data.

In this study we noted that the normalised residuals between the posterior mean and observed sky have a mean of -0.23 and standard deviation of 0.83. The standard deviation being smaller than 1 may indicate that the reported observational uncertainty maps (or assumed 10\% when no uncertainty map is reported) for the observations are too large. We could potentially address this by fitting for the noise level in the observations. To do this we would assume that the noise in the observations is some fraction of the observed temperature, such that the noise map $N_v(\Omega)$ at frequency $v$ is given as $N_v(\Omega) = \alpha_v D_v(\Omega)$. The fractional noise term for each observed frequency $\alpha_v$ could then be infered as an extra model parameter. This would avoid us having to trust that the reported uncertainty maps are correct, and it would avoid us having to assume an arbitrary 10\% noise level for frequencies that do not have reported uncertainty maps. 

In this study bright point sources are masked out and are not modelled. However, previous studies have shown that contamination from point sources is sufficient to cause a systematic bias in 21-cm signal recovery~\citep{Mittal2024}. In future versions of B-GSM we will aim to include point sources in our foreground model. We could potentially model the point source contribution as a third emission component with a parametrised spectral model. This could then be included into the existing B-GSM framework. Additionally, we could investigate the inclusion of independent point source datasets such as the GLEAM low-frequency extragalactic catalogue~\citep{Hurley-Walker2016} into the inference to improve modelling of point sources in a future version of B-GSM.

In addition to improved noise modelling and calibration, a future aim is to explore more general spectral models. In this study we restricted ourselves to only considering curved power-law spectra for the emission components. However this restriction is not fundamental to B-GSM which can handle any spectral model. In future work we aim to explore non-parametric spectral models such as FlexKnots \citep{Shen2024}. 

A more long term research goal is to refine B-GSMs noise model to better account for correlated noise in the diffuse maps. This could potentially improve the accuracy of the posterior sky predictions, and better account for the structure and non-Gaussian distribution seen for the normalised residuals in this study. Accounting, for correlated noise is difficult as we will no longer be able to treat pixels as statistically independent. This will result in a significant increase in computational complexity, as the likelihood would be defined on a map-by-map level requiring inversion of very large $(k\cdot n_\mathrm{pix})\times(k\cdot n_\mathrm{pix})$ matrices. Potential future research could explore the use of Simulation Based Inference techniques e.g. normalising flows, to accelerate this inference and avoid the assumptions that we made when defining our likelihood function in this study. 

In this study we smooth all the diffuse maps to a common resolution of $5\degr$. This smoothing during pre-processing throws away a large amount of high resolution information, for example the Haslam 408~MHz map has a native beam size of 56 arcmin \citep{Remazeilles2015}. Future work could explore inference at the native resolution of the diffuse maps. In essence we would use our component separation model to predict the sky at each of the observed frequencies at the resolution of the highest resolution map. These predicted maps would then be convolved with a beam model to smooth them to the native resolution of the observed map at each frequency, this would then allow comparison of the predicted and observed sky and calculation of a likelihood. This would require convolutions of the predicted sky at each frequency with a corresponding beam model. To do this we would need accurate models of the beams for all diffuse maps used in this study. The convolutions would greatly increase the computational cost of likelihood evaluations and would not allow us to assume that pixels are statistically independent. As with accounting for correlations in the noise, this proposed future research direction could potentially be approached using SBI techniques with the convolutions occurring as part of the simulator.
%A more long term goal is to apply the B-GSM framework for fully Bayesian Global Signal extraction with a physically motivated foreground model.
%%%%%%%%%%%%%%%%%%%%%%%%%%%%%%%%%%%%%%%%%%%%%%%%%%
\section*{Data Availability}
The absolute temperature data used in this study is taken from the figures in the papers by \cite{Mozdzen2016}, \cite{Mozdzen2018} and \cite{Monsalve2021}. The CSV files containing the digitised temperatures from these figures are available from the GitHub repository, along with the pre-processed diffuse dataset used for this study. All code and the posterior samples for B-GSM (including code for producing the plots in this paper) is available for public download from the following GitHub repository:\\ \url{https://github.com/George-GTC30/Bayesian-Global-Sky-Model-B-GSM-Paper-2}

%%%%%%%%%%%%%%%%%%%% REFERENCES %%%%%%%%%%%%%%%%%%

% The best way to enter references is to use BibTeX:

\bibliographystyle{mnras}
\bibliography{example} % if your bibtex file is called example.bib

%%%%%%%%%%%%%%%%%%%%%%%%%%%%%%%%%%%%%%%%%%%%%%%%%%

% Don't change these lines
\bsp	% typesetting comment
\label{lastpage}
\end{document}